\documentclass[cits]{PoS}
\usepackage{amsmath,amssymb}
\newcommand{\Tr}[0]{\text{Tr}}

\title{Block renormalization group transformations and overlap fermions}

\ShortTitle{Block renormalization group transformations and overlap fermions}

\author{\speaker{Nigel Cundy}\\
        Universit\"at Regensburg, Universit\"at Stra\ss e 31, 93040 Regensburg\\
        E-mail: \email{nigel.cundy@physik.uni-regensburg.de}}

\abstract{In this preliminary work, I provide the outline of an argument (leaving the full proof to a future publication) that there exists a valid renormalization group blocking transformation which converts the continuum fermion action into a Ginsparg-Wilson lattice action. I construct the blocking for the massless overlap operator as a specific example, indicating how other Ginsparg-Wilson lattice Dirac operators can be derived in a similar fashion. This renormalization group transformation modifies the gauge action and adds a number of irrelevant terms to the lattice action. The procedure is not valid for lattice Dirac operators which do not exactly satisfy the Ginsparg-Wilson relation, for example the Wilson operator.}

\FullConference{The XXVI International Symposium on Lattice Field Theory\\
		 July 14-19 2008\\
		 Williamsburg, Virginia, USA}

\begin{document}

\section{Introduction}
Since chiral symmetry determines many of the low energy properties of QCD, lattice simulations should ideally respect chiral symmetry to a high accuracy. The Ginsparg-Wilson framework~\cite{Ginsparg:1982bj,Luscher:1998pqa} offers a method of constructing lattice Dirac operators with an exact `chiral symmetry.' Examples of Ginsparg-Wilson lattice Dirac operators have been known for some time, and the overlap operator, which exactly satisfies the Ginsparg-Wilson chiral symmetry, is currently being used in lattice simulations of two flavour~\cite{Cundy:2005pi} and 2+1 flavour~\cite{Hashimoto:2007vv} lattice QCD. 

However, it is long been known that the overlap~\cite{Narayanan:1993sk} and perfect action~\cite{Hasenfratz:1994,Bietenholz:1995nk} fermions are not the only possible Ginsparg-Wilson Dirac operators, and there is some theoretical interest in trying to describe the connection between the various possible solutions to the Ginsparg-Wilson equation. In~\cite{cundy:2008cs}, I described the derivation of a group of valid chiral lattice Dirac operators, based on the overlap operator, all of which satisfy a Ginsparg-Wilson relation. The general form for massless Dirac operators of this type was given as
\begin{gather}
D_N' = \frac{1}{{s}(\gamma_5\epsilon)}{t}\left[\frac{1}{2}(\gamma_5\epsilon + \epsilon \gamma_5)\right]{h}\left({q}\left[\frac{1}{2}(\gamma_5\epsilon + \epsilon \gamma_5)\right]\left(1 + {r} \left[\frac{1}{2}(\gamma_5\epsilon + \epsilon \gamma_5)\right]\gamma_5\epsilon\right)\right),\label{eq:1}
\end{gather}
where $s$, $t$, $h$, $q$ and $r$ are analytic functions, only constrained by the need to preserve the correct continuum limit and give the doublers an infinite mass, and $\epsilon = \text{sign}(\gamma_5 D_W)$, where $D_W$ is a kernel operator, and, as with overlap fermions, can be any valid lattice Dirac operator at a negative mass (for example the Wilson operator).

It remains unlikely that any of these operators offer advantages over standard overlap fermions. However, there is one intriguing case, where $r[x]=1/|x|$ (neglecting that proof of the locality of the Dirac operator given in~\cite{cundy:2008cs} depends on the analyticity of $r$; although that the proof breaks down does not necessarily mean that the operator itself is non-local), with $q$ arbitrary, and the other functions equal to one. This Dirac operator has an eigenvalue spectrum on two parallel lines, the physical modes along the imaginary axis, and the doublers at infinite mass: mirroring the properties of the continuum operator. With the ability to use the function $q$ to arbitrarily place the eigenvalues along the imaginary axis (except for the zero modes), this offers the tantalising possibility to construct a lattice Dirac operator whose eigenvalue spectrum precisely matches that of the continuum (provided that the lattice spacing is sufficiently fine that the lattice topological index matches the continuum topological charge). Since almost every observable in lattice QCD can be expressed in terms of the overlap operator~\cite{Alexandru:2008fu}, and hence its eigenvalue spectrum, this operator would provide an intriguing possibility to reduce the artefacts of a continuum simulation: it could be a candidate perfect action. Of course, finding the crucial function $q$ remained and remains elusive; and until it is found the hypothesis that such a perfect action can be constructed in this way remains questionable.  

Assuming that such a perfect action exists, a question arises concerning whether there is a connection between this purported `perfect action' and the long-established perfect action derived from renormalization group considerations~\cite{Hasenfratz:1994}. This naturally leads to an examination of the place of overlap fermions within the language of the renormalization group. In this very preliminary study, I start an investigation of the place of overlap-style lattice Dirac operators within  the renormalization group, in the hope of better understanding their theoretical foundation. These proceedings should thus only be percieved as a few brief ruminations while I leave a more detailed construction to a future work~\cite{cundyforthcoming}. In section \ref{sec:2}, I outline the required renormalization group tools; in section \ref{sec:3} I apply these to the case of overlap fermions on the lattice; and in section \ref{sec:4} I use these results to give an alternative derivation of equation (\ref{eq:1}).
\section{Block renormalization transformations}\label{sec:2}
I can define a block transformation, used to transform a spinor $\psi_0$ and Dirac operator $D_0$ to a spinor $\psi_1$ and Dirac operator $D_1$ in terms of three functions $\alpha$, $B$, and $B_d$:
\begin{align}
Z=\int& d\overline{\psi}_0d\psi_0e^{-\overline{\psi}_0D_0\psi_0 - \frac{1}{g_0^2}F_{\mu\nu}^2} = N({\alpha, \beta})\int d\psi_1 d\overline{\psi}_1 \int d\overline{\psi}_0d\psi_0e^{-\overline{\psi}_0(x) D_0 \psi_0(x)-\frac{1}{{g_0}^2}F_{\mu\nu}^2} \times\nonumber\\
\phantom{sospace}& e^{-(\overline{\psi}_1(x) - {\overline{\psi}_0}(x') {B^{-1}_d}(x',x)){\alpha}(x,y)  ({\psi_1(y)} - {B}^{-1}(y,y'){\psi_0(y')})} 
= \int d{\psi_1} d{\overline{\psi}_1}e^{-{\overline{\psi}_1(x) D_1 \psi_1(x)}-\frac{1}{{g_1}^2}F_{\mu\nu}^2},\label{eq:2.1}
\end{align}
where $N$ is a normalisation factor, and I will always assume integration (in the continuum) or summation (on the lattice) over repeated indices. I will also assume that $B$, $B_d$ and $\alpha$ are all analytic and invertible (in some sense; precisely what I mean by `invertible' lies beyond the scope of this work). The derivation of the last equality only requires that  $B_d^{-1}\alpha B^{-1}$  is positive-definite, that $B_d^{-1}\alpha B^{-1} +D_0$ is invertible and that the `inverse' blockings $B$ and $B_d$ exist.  

Suppose that the original fermion action, $\overline{\psi}_0 D_0 \psi_0$ is invariant under an infinitesimal chiral rotation:
\begin{align}
\overline{\psi}_0 \rightarrow &\overline{\psi}_0 e^{i{\epsilon} \gamma_5}; & \psi_0 \rightarrow& e^ {i{\epsilon} \gamma_5}\psi_0.
\end{align}
Following the methods of Ginsparg and Wilson~\cite{Ginsparg:1982bj} and L\"uscher~\cite{Luscher:1998pqa}, it can easily be shown that this implies a symmetry of the transformed action, where the new Dirac operator obeys the modified Ginsparg-Wilson relation
\begin{gather}
D_1 B^{-1} \gamma_5 B + B_d \gamma_5 B_d^{-1} D_1 = D_1(\alpha^{-1}B_d\gamma_5 B_d^{-1} + B^{-1}\gamma_5 B\alpha^{-1})D_1,\label{eq:modGW}
\end{gather}
and the corresponding `chiral symmetry' transformations of the spinor fields are
\begin{align}
\psi \rightarrow& e^{i\epsilon \gamma_5(\gamma_5B^{-1} \gamma_5 B - \gamma_5B^{-1} \gamma_5 B\alpha^{-1}D_1)} \psi&
\overline{\psi} \rightarrow &\overline{\psi} e^{i\epsilon (B_d \gamma_5 B_d^{-1}\gamma_5 -D_1\alpha^{-1}B_d \gamma_5 B_d^{-1}\gamma_5)\gamma_5},
\end{align}
while the topological charge is
\begin{gather}
Q_f = \frac{1}{2}\Tr (B^{-1} \gamma_5 B + B_d \gamma_5 B_d^{-1} - B^{-1} \gamma_5 B\alpha^{-1}D - D\alpha^{-1}B_d \gamma_5 B_d^{-1} )
\end{gather}
If $B$ and $B_d$ commute with $\gamma_5$, which both Ginsparg and Wilson's original work~\cite{Ginsparg:1982bj} and the construction of the standard forms of the fixed point action~\cite{Hasenfratz:1994} assume then, unsurprisingly, equation (\ref{eq:modGW}) reduces to the canonical Ginsparg-Wilson relations. Here I consider the different case where $\alpha$ is proportional to the unit matrix and has a value of $\infty$ and $B$ does not commute with $\gamma_5$. When $\alpha=\infty$, the integral over $\psi_0$ in equation (\ref{eq:2.1}) can be performed explicitly, giving a `Jacobian' $e^{-J} = \det (B_d^{-1}B^{-1})$ (together with an unimportant constant factor) and a new Dirac operator $D_1 = B_d D_0 B$. Of course, on the lattice, for this chiral symmetry to reduce to the correct continuum expression, $\gamma_5B \gamma_5 B^{-1}$ and $\alpha$ should exist, be local, invertible, and reduce to $1+O(a)$ and $\infty$ respectively.  

It is possible to construct a blocking transformation from the eigenvectors of the Dirac operator. I introduce a large negative mass $\Lambda$, which will act as a momentum cut-off. For the continuum Dirac operator, the eigenvalue equations (for each pair of non-zero eigenvalues) can be written as
\begin{align}
(D_0 + \Lambda) \phi_+(x,\lambda) =& (\Lambda + i\lambda)\phi_+(x,\lambda)\\
(D_0 + \Lambda) \phi_-(x,\lambda) =& (\Lambda - i\lambda)\phi_-(x,\lambda)\\
\phi_+(x,\lambda) = \gamma_5 \phi_-(x,\lambda)
\end{align}
From these eigenvalues and eigenfunctions, we can construct the eigenfunctions and eigenvalues of the Hermitian Dirac operator $\gamma_5 (D_0+\Lambda)$. The eigenfunctions are
\begin{gather}
\frac{1}{\sqrt{2}}\left(\phi_+(x,\lambda) \pm e^{i\eta(\lambda,\Lambda)}\phi_-(x,\lambda)\right),
\end{gather}
with eigenvalues $\pm \mu$ where
\begin{align}
\mu = &\sqrt{\Lambda^2 + \lambda^2}& e^{i\eta(\lambda,\Lambda)} = & \frac{i\lambda + \Lambda}{\sqrt{\Lambda^2 + \lambda^2}}.
\end{align}
We can use this formulation to construct a renormalization group blocking, with $\alpha=\infty$ and
\begin{align}
B =& B_d = \sum_{\text{zero modes}} \phi_0(x',0)\phi_0^{\dagger} + \int d\lambda\rho(\lambda)\left[\phi_+(x,\lambda)\phi_+^{\dagger}(x',\lambda) H\left( \left[\Lambda \frac{1+e^{i\eta(\lambda,\Lambda)}}{i\lambda}\right]^{\frac{1}{2}}\right) +\phantom{a}\right.\nonumber\\&\phantom{spaceandspaceandspaceandspace}\left. \phi_-(x,\lambda)\phi_-^{\dagger}(x',\lambda) H\left( \left[\Lambda \frac{1+e^{i\eta(-\lambda,\Lambda)}}{-i\lambda}\right]^{\frac{1}{2}}\right)\right],
\end{align}
where $\rho$ is the density of eigenvalues and $H$ is an otherwise arbitrary analytic function satisfying $\Re(H[z]^2) > 0$ for $\Re[z] > 0$, which, given that  $\Re((1+e^{i\eta})/\lambda) > 0$ ensures that the condition that the real part of the eigenvalues of $B_d^{-1} B^{-1}$  must be positive is satisfied. For example, taking $H=1$ gives a new continuum Dirac operator
\begin{gather}
D_1 = 1+\gamma_5\text{sign}(\gamma_5(D_0 + \Lambda)),
\end{gather}
where the eigenvalues of this continuum Dirac operator $D_1$ lie on a semi-circle of radius 1 in the complex plane. 
The Jacobian for this blocking is
\begin{gather}
J=\int_{\lambda\neq0} \rho(\lambda)\log\left[1 + \frac{\Lambda}{\sqrt{\Lambda^2 + \lambda^2}} \frac{2 \Lambda^2}{\lambda^2}\right].
\end{gather}
Assuming that we can regulate the Dirac operator so that we can neglect those eigenvalues which are not significantly smaller than $\Lambda$, we can expand $J$ in powers of $D_0^{\dagger}D/\Lambda$, which gives
\begin{gather}
J = \Tr\left[\frac{3}{\Lambda^2} D^{\dagger}_{\mu}D_{\mu} - \frac{29}{8\Lambda^4}(D^{\dagger}_{\mu}D_{\mu})^2 \right]- \frac{29}{8\Lambda^4}\sum_x F_{\mu\nu}^2(x) + O(\Lambda^{-6}).
\end{gather}
When I apply these methods to the lattice, assuming that $D_0$ is an ultra-local Dirac operator, $\Tr (D^{\dagger}_{\mu}D_{\mu})$ will be a constant (independent of the gauge field),  and this Jacobian just gives a $\beta-$shift, as implied in equation (\ref{eq:2.1}). In the continuum, these trace terms can be re-expressed in terms of a bosonic field. 

Of course, $D_1$ is only exponentially local, with a rate of decay controlled by $\Lambda$, and if we insist that in the continuum the Dirac operator must be `ultra-local,' then this transformation would only be valid in the limit that $\Lambda\rightarrow\infty$, where we recover the original operator $D_0$. On the lattice, this situation is tolerable and controlable as long as $\Lambda$ is inversely proportional to the lattice spacing. This exponential locality will always be seen if $H$ is analytic and $\Lambda$ negative~\cite{cundyforthcoming}. This implies that the eigenvalues of the new Dirac operator will lie on a closed loop in the complex plane. In the general case, it is clear that if $D_0$, $B_d$ and $B$ are local, then $D_1 = B_dD_0B$ will also be local.
\section{The renormalization group and overlap fermions}\label{sec:3}
To move from the continuum to the lattice, one needs a renormalization group blocking which averages (in some respect) over the continuum fields. I introduce a blocking procedure $P^C_L(l,x)$, a function of the continuum gauge fields, so that the lattice fermion fields $\psi_L(l)$ can be generated from the continuum fields $\psi_C(x)$ by the relation $\psi_L(l) = \int d^4x P^C_L(l,x)\psi_C(x)$. It is also desirable to have an inverse blocking $\psi_C(x)=\sum_lP_L^C(x,l)\psi_L(l)$, so that the continuum theory can be extracted from the lattice theory by a continuous renormalization group transformation. These blockings have to be local, and include a means of removing the doublers in the corresponding Dirac operator; but otherwise we are very free in our choice of the blocking. Finding a blocking to move from the continuum to a particular lattice discretization is straight-forward. Finding a suitable inverse, to reach the continuum from the lattice, is not. For example, it is `possible' to generate a Wilson-type fermion action from continuum fermions using the blocking
\begin{align}
\psi_C(y) =& \sum_x \prod_{\beta} \theta\left(|l_{\beta}-y_{\beta}|-\frac{a}{2}\right)\left(1+\sum_{\alpha}\gamma_{\alpha}n(l_{\alpha}-y_{\alpha})\right)e^{-i\int^l_yA_{\mu}(y')dy'_{\mu}}\psi_L(l) = \sum_x B_W(y,l)\psi_L(l)\nonumber\\
\overline{\psi}_C(y)=&\sum_l \overline{\psi}_L(l)\prod_{\beta} \theta\left(|l_{\beta}-y_{\beta}|-\frac{a}{2}\right)\left(1-\sum_{\alpha}\gamma_{\alpha}n(l_{\alpha}-y_{\alpha})\right)e^{i\int^l_yA_{\mu}(y')dy'_{\mu}}= \sum_l \overline{\psi}_L(l)B_{W,d}(l,y)\label{eq:3.1}
\end{align}
where $n$ is the function
\begin{gather}
n(x)=\left\{\begin{array}{r l}
1&\;\;\;x>(1-|\epsilon|)a/2\\
0&\;\;\; |x|<(1-|\epsilon|)a/2\\
-1 &\;\;\; x<(1-|\epsilon|)a/2
\end{array}
  \right.
\end{gather}
in the limit that $\epsilon\rightarrow0$. Inserting (\ref{eq:3.1}) into the continuum action, $\overline{\psi}_C(y') D_C(y',y) \psi_C(y)$, gives an action $\overline{\psi}_L(l') D_W(l',l)\psi_L(l)$. The only difference between this action and the standard Wilson fermion action is how the link $U$ is constructed from the the gauge field $A_{\mu}$: rather than a single parallel transporter, this new `link' is a sum of parallel transporters. These blockings can be used as a basis for the transformations needed to generate overlap fermions. As an example,
\begin{align}
B_{O,d} =& B_{W,d};&
B_O =& D_W^{-1} (1+\gamma_5\phi_W\phi^{\dagger}_W\text{sign}(\lambda_W))B_W,
\end{align}
where $\phi_W$ and $\lambda_W$ are the eigenvectors and eigenvalues of $\gamma_5 (D_W+\Lambda)$, generates the overlap action. 

However, the Wilson-fermion `blocking' given in equation (\ref{eq:3.1}) is not valid, because although $ B_d D_C B$ gives an object similar to the Wilson operator, the matrix $B_{W,d}^{-1} B_W^{-1}$ is not positive definite; therefore, as mentioned earlier, the last equality of equation (\ref{eq:1}) is invalid. However, for overlap fermions  $B_{O,d}^{-1} B_O^{-1}$ is positive definite (provided that the lattice spacing is sufficiently fine); thus this is a valid renormalization group transformation. I shall leave the formal proof of this statement to the subsequent publication~\cite{cundyforthcoming};  for now I will simply say that it is strongly suggested by the well-known fact that overlap fermions satisfy the Ginsparg-Wilson relation while Wilson fermions do not.

There are alternative formalisms which can be used to construct the $\alpha$ and $B$ blocking functions to obtain a particular lattice Dirac operator $D_L$, given some transformation $P_L^C$ which averages over the continuum fields in a suitable manner to obtain the lattice fields. One possibility is the symmetric choice $B_d=D_L^{1/2} P_L^C D_C^{-1/2}$, $B = D_C^{-1/2}\overline{P}_L^C D_L^{1/2}$, and $\alpha=\infty$, which leads to a Ginsparg-Wilson relation
\begin{gather}
0 = D_L\gamma_5\hat{\gamma}_5\gamma_5 + \hat{\gamma}_5 D_L,\label{eq:3.3}
\end{gather}
where
\begin{gather}
\hat{\gamma}_5 = \frac{D_L}{\sqrt{ D_L^{\dagger} D_L}} \gamma_4.
\end{gather}
It is easy to demonstrate that equation \ref{eq:3.3} is satisfied if we replace $D_L$ with the overlap operator. The Jacobian of the blocking is
\begin{gather}
J = \Tr \left[\log\left(D_C\right) - \log(D_L) - \log (P^C_L \overline{P}_L^C)\right] = -\frac{1}{4g_R^2}a^4 F_{\mu\nu}^2 + \ldots,
\end{gather}
and again this transformation preserves the action, just giving a shift in the coupling constant.
This choice of $B$ is not unique~\cite{Mandula:2007jt}; but the work here suggests that the different ways of expressing the chiral symmetry on the lattice are related to different lattice regularisations (since each of these possible blockings will give a different shift to the gauge coupling). For example, if we choose $B_d = \overline{P}_L^C$, $B = D_C^{-1} P_L^C D_L$, we recover the canonical Ginsparg-Wilson relation. It is to be stressed that unlike Ginsparg and Wilson's original chiral fermions, and perfect action fermions which use a different blocking with a finite $\alpha$, the lattice chiral fermions discussed here are generated with $\alpha = \infty$ and a blocking field $B$ which does not commute with $\gamma_5$.  

\section{New Solutions to the Ginsparg-Wilson operator}\label{sec:4}
Once we have found one chiral lattice Dirac operator, it is easy to generate more by constructing blockings from the eigenvalues of the operator. For example, by witing the overlap eigenvalues as $1+e^{i\theta}$ with eigenvectors $\Theta$, we can construct a blocking 
\begin{align}
\psi_L(y) = &\sum_x \sum_\theta F(\theta) \Theta(\theta,y)\Theta^{\dagger}(\theta,x)\psi'_L(x)\nonumber\\
\overline{\psi}_L(y) = &\sum_x\overline{\psi}'_L(x) \sum_\theta F(\theta) \Theta^{\dagger}(\theta,x)\Theta(\theta,y)\nonumber\\
F=&\sqrt{\frac{1}{1+e^{i\theta}} \frac{1}{s(e^{i\theta})} t(\cos\theta)h[q(\cos\theta)\left(1+r(\cos\theta)e^{i\theta}\right)]}
\end{align}
where $s$, $t$, $h$, $q$ and $r$ are analytic functions (chosen to ensure the correct continuum limit and that the fermion doublers have infinite mass in the continuum, which can be achieved by ensuring that $F$ is analytic, positive, and that $F=1$ at $\theta = 0,\pi,2\pi$), which leads to the generalised overlap operator given in equation (\ref{eq:1}).
That this operator is exponentially local is clear from the analyticity of the momentum representation of the renormalization group transformation, the exponential locality of the overlap operator and that the overlap operator commutes with its Hermitian conjugate. 

\section{Conclusions}\label{sec:5}
I have tentatively suggested that there is, in an infinite volume, a renormalization group transformation linking the continuum fermion action with a Ginsparg-Wilson lattice fermion action, and I have shown how such a blocking can be constructed for massless overlap fermions. However, a different style of block transformation is used compared to that in  fixed point fermions or original construction of the Ginsparg-Wilson relation. The Jacobian from the block transformation corresponds to a shift in the gauge coupling, together with some irrelevant terms; thus the lattice QCD Lagrangian is preserved by this transformation. 

Clearly there is one crucial step missing in the argument as outlined here, namely a proof that for the overlap operator the matrix, defined here as $B_d^{-1}B^{-1}$, is positive definite. I shall attempt to provide that proof in a future publication~\cite{cundyforthcoming}.
However, if this conclusion is correct, then it could have implications in the analysis of discretization errors, and in the theoretical description of a perfect action derived from overlap fermions.

\bibliographystyle{JHEP}
\bibliography{new_GW_operator}

\providecommand{\href}[2]{#2}\begingroup\raggedright\begin{thebibliography}{10}

\bibitem{Ginsparg:1982bj}
P.~H. Ginsparg and K.~G. Wilson, {\it A remnant of chiral symetry on the
  lattice},  {\em Phys. Rev.} {\bf D25} (1982) 2649.

\bibitem{Luscher:1998pqa}
M.~L{\"u}scher, {\it Exact chiral symmetry on the lattice and the ginsparg-
  wilson relation},  {\em Phys. Lett.} {\bf B428} (1998) 342--345,
  [\href{http://xxx.lanl.gov/abs/hep-lat/9802011}{{\tt hep-lat/9802011}}].

\bibitem{Cundy:2005pi}
N.~Cundy {\em et~al.}, {\it Numerical methods for the {QCD} overlap operator.
  {IV}: Hybrid monte carlo},  {\em Computer Physics Communications} (2008)
  [\href{http://xxx.lanl.gov/abs/hep-lat/0502007}{{\tt hep-lat/0502007}}].

\bibitem{Hashimoto:2007vv}
{\bf JLQCD} Collaboration, S.~Hashimoto {\em et~al.}, {\it {Lattice simulation
  of 2+1 flavors of overlap light quarks}},  {\em PoS} {\bf LAT2007} (2007)
  101, [\href{http://xxx.lanl.gov/abs/hep-lat/0710.2730}{{\tt
  hep-lat/0710.2730}}].

\bibitem{Narayanan:1993sk}
R.~Narayanan and H.~Neuberger, {\it Chiral determinant as an overlap of two
  vacua},  {\em Nucl. Phys.} {\bf B412} (1994) 574--606,
  [\href{http://xxx.lanl.gov/abs/hep-lat/9307006}{{\tt hep-lat/9307006}}].

\bibitem{Hasenfratz:1994}
P.~Hasenfratz and F.~Niedermayer {\em Nucl. Phys.} {\bf B414} (1994) 785.

\bibitem{Bietenholz:1995nk}
W.~Bietenholz and U.~J. Wiese, {\it {A Perturbative construction of lattice
  chiral fermions}},  {\em Phys. Lett.} {\bf B378} (1996) 222--226,
  [\href{http://xxx.lanl.gov/abs/hep-lat/9503022}{{\tt hep-lat/9503022}}].

\bibitem{cundy:2008cs}
N.~Cundy, {\it {New solutions to the Ginsparg-Wilson equation}},  {\em Nucl.
  Phys.} {\bf B802} (2008) 92--105,
  [\href{http://xxx.lanl.gov/abs/hep-lat/0802.0170}{{\tt hep-lat/0802.0170}}].

\bibitem{Alexandru:2008fu}
A.~Alexandru, I.~Horvath, and K.-F. Liu, {\it {Classical Limits of Scalar and
  Tensor Gauge Operators Based on the Overlap Dirac Matrix}},
  \href{http://xxx.lanl.gov/abs/hep-lat/0803.2744}{{\tt hep-lat/0803.2744}}.

\bibitem{cundyforthcoming}
N.~Cundy.
\newblock In Preparation.

\bibitem{Mandula:2007jt}
J.~E. Mandula, {\it {Note on the Lattice Fermion Chiral Symmetry Group}},
  \href{http://xxx.lanl.gov/abs/hep-lat/0712.0651}{{\tt hep-lat/0712.0651}}.

\end{thebibliography}\endgroup

\end{document}